\documentstyle[emulateapj]{article}

\received{}
\revised{}
\accepted{}
\journalid{}{}
\articleid{}{}
\paperid{}
\cpright{}{}
\ccc{}

\slugcomment{Submitted to the Astrophysical Journal}

\lefthead{Wood, Wolff, Bjorkman, \& Whitney}
\righthead{Modeling The Spectral Energy Distribution of HH30~IRS}

\begin{document}

\title{The Spectral Energy Distribution of HH30~IRS: \\
Constraining The Circumstellar Dust Size Distribution}

\author{Kenneth Wood\altaffilmark{1,2}, Michael~J.~Wolff\altaffilmark{3}, 
J.E.~Bjorkman\altaffilmark{4}, Barbara Whitney\altaffilmark{3}}

\altaffiltext{1}{School of Physics \& Astronomy, University of St Andrews, 
North Haugh, St Andrews, Kingdom of Fife, KY16 9AD, Scotland the Brave; 
kw25@st-andrews.ac.uk}

\altaffiltext{2}{Harvard-Smithsonian Center for Astrophysics, 
60 Garden Street, Cambridge, MA~02138; kwood@cfa.harvard.edu}

\altaffiltext{3}{Space Science Institute, 3100 Marine Street, Suite A353, 
Boulder, CO~80303; bwhitney@colorado.edu, wolff@colorado.edu}

\altaffiltext{4}{Ritter Observatory, Department of Physics \& Astronomy, 
University of Toledo, Toledo, OH 43606; jon@astro.utoledo.edu}

\authoremail{kw25@st-andrews.ac.uk}

\begin{abstract}

We present spectral energy distribution (SED) models for the edge-on 
classical T~Tauri star HH30~IRS that indicate dust grains have grown to 
larger than $50\mu{\rm m}$ within its circumstellar disk.  
The disk geometry and inclination are known from previous modeling of 
multiwavelength {\it Hubble Space Telescope} images and we 
use the SED ($0.5\mu {\rm m} \le \lambda \le 3$mm) to constrain the dust size 
distribution.  Model spectra are shown for different circumstellar 
dust models: a standard ISM mixture and larger grain models.    
As compared to ISM grains, the larger dust grain models have a shallower 
wavelength dependent opacity: smaller at short 
wavelengths and larger at long wavelengths.  
Models with the larger dust grains provide a good 
match to the observed SED of HH30~IRS.  Although the currently available 
SED is poorly sampled, we estimate $L_\star \approx 0.2L_\odot$, 
$M_{\rm disk} \approx 1.5\times 10^{-3}M_\odot$, and a power-law with
exponential cutoff dust grain size 
distribution.  This model provides a good fit to the currently 
available data, but mid and far-IR observations are required to more 
tightly constrain the size distribution.  
The accretion luminosity in our models is 
$L_{\rm acc}\la 0.2 L_\star$ corresponding to an accretion rate 
$\dot M \la 4 \times 10^{-9}M_\odot {\rm yr}^{-1}$.  
Dust size distributions 
that are simple power-law extensions (i.e., no exponential cutoff) 
yield acceptable fits to the optical/near-IR but too 
much emission at mm wavelengths and require larger 
disk masses up to $M_{\rm disk}\sim 0.5M_\odot$.  Such a simple
size distribution would not be expected in an environment such as the
disk of HH30~IRS, (i.e., where coagulation and accretion processes are
occurring in addition to grain shattering), particularly over such a
large range in grain sizes.  However, its ability to adequately characterize
the grain populations may be determined from more complete observational
sampling of the SED in the mid to far-IR.

\end{abstract}


\section{Introduction}

The {\it HST} WFPC2 images of HH30~IRS (Burrows et al. 1996) revealed 
the characteristic scattered light pattern of an edge-on disk in a classical 
T~Tauri system: two almost parallel nebulae separated by an opaque dust lane 
(e.g., Whitney \& Hartmann 1992).  Detailed modeling of the 
WFPC2 images enabled Burrows et al. to constrain the size, shape, and optical 
depth of the disk.  
In addition, their best model images were constructed using 
dust grains with a scattering phase function which is more forward throwing 
than typical interstellar medium (ISM) grains, indicative of grain growth 
within the disk (Hansen \& Travis 1974).  

Wood et al. (1998) presented scattered light models of edge-on disks 
showing that the width of the dust lane decreases towards 
longer wavelengths where the dust is less opaque (assuming a wavelength 
dependence for absorption and scattering of typical ISM dust grains).  They 
predicted that the HH30~IRS nebulosity should appear almost pointlike at the 
$K$ band.  However, the new NICMOS observations of Cotera et al. (2001) 
show a distinct dust lane even at the longest NICMOS wavelength.  
Cotera et al. used the NICMOS images to study the near-IR wavelength 
dependence of the circumstellar dust opacity.  They modeled the NICMOS 
images using a dust 
size distribution that has a smaller near-IR opacity and a shallower 
wavelength dependence than ISM dust.  Their modeling showed that the 
extinction through the disk to the central source is $A_V = 7900$.  
The derivation of this extinction requires detailed radiation transfer 
modeling, since the {\it HST} images are dominated by scattered light so 
standard reddenning estimates are not applicable.

The scattered light models of Burrows et al., Wood et al. (1998), and 
Cotera et al. fit the width of the dust lane, from which they derive 
$M_{\rm disk}\ga 10^{-4}M_\odot$.  
As noted by Burrows et al., what is really determined from scattered light 
modeling is the equatorial optical depth in the disk midplane, 
$\tau_{eq}\propto\kappa_{\lambda} M_{\rm disk}$.  
The mass determination is therefore very 
uncertain since it depends on the adopted grain size distribution 
and opacity.  Scattered light models of the edge-on disk system HK~Tauri/c 
derive $M_{\rm disk} \approx 10^{-4} M_\odot$ assuming ISM dust 
(Stapelfeldt et al. 1998).  However, D'Alessio, Calvet, \& Hartmann (2001) 
have shown that the edge-on disk of HK~Tau/c may be modeled with a massive 
disk, $M_{\rm disk} = 0.065 M_\odot$, 
in which the maximum grain size is 1~meter.  Their results also
demonstrate the utility of long wavelength observations for distinguishing
between a low mass disk with ISM grains and a massive disk with a population
of much larger grains.  

It is difficult to draw firm conclusions on circumstellar dust 
properties since the often unknown inclination 
and disk geometry lead to degeneracies in SED models.  However, for 
HH30~IRS scattered light modeling provides the disk geometry and 
inclination.  Therefore the circumstellar grain properties may be studied 
through detailed modeling of its SED which covers the range 
$0.55\mu{\rm m}\le \lambda\le 3$mm (Stapelfeldt \& Moneti 1999).  
This paper presents model SEDs for HH30~IRS that highlight 
the differences between 
ISM and larger dust grain models.  Our modeling adopts the disk structure 
derived from fitting {\it HST} images and calculates the radiative 
equilibrium SED from a star plus disk system.  We use 
our new Monte Carlo radiative equilibrium code for the calculations and 
show SED models at different inclination angles ranging from almost edge-on 
(appropriate to HH30~IRS) to pole-on viewing of the star~+~disk system.  

Section~2 presents the HH30~IRS SED taken from the literature, \S3 describes 
the radiation transfer, disk model, and dust opacity, and \S4 
presents SED models for a passive reprocessing disk and also the effects of 
including viscous accretion luminosity.  \S5 summarizes our results.

\section{The Spectral Energy Distribution of HH30~IRS}

Our models are constructed with the goal of reproducing the currently 
available SED for HH30~IRS.  
The SED displayed in Fig.~1 is taken from Mundt \& Fried (1983: $V$ and $I$), 
Vrba, Rydgren, \& Zak (1985: $H$, $K$, and an upper limit at $L$), and 
Reipurth et al. (1993: 1.3mm).  The SED 
measured by {\it ISO} between $4.5\mu$m and $90\mu$m and additional mm 
fluxes (Stapelfeldt \& Padgett 2001) were presented by 
Stapelfeldt \& Moneti (1999, Fig.~3). 

The {\it ISO} data were also 
presented by Brandner et al. (2000) and their reduction produced 
slightly larger fluxes than that of Stapelfeldt \& Moneti (1999).  
Both reductions are shown on this figure.   The fluxes in Fig.~1 were not 
taken simultaneously and HH30~IRS is known to vary in the optical by up 
to 1.4mag at $V$, but this amplitude decreases towards longer wavelengths 
(Wood et al. 2000).  Consequently, this variability will not affect our 
dust size distribution modeling since the different dust models 
are discriminated at long wavelengths ($\lambda\ga 20\mu{\rm m}$) where 
any variability is expected to be small.

Although the $60\mu$m and $90\mu$m data are upper limits, the SED shows 
the characteristic double-peaked spectrum of an edge-on disk: the direct 
starlight is obscured by the disk, the near-IR peak arises from scattered 
light, and the far-IR peak is due to reprocessing of stellar radiation in the 
disk and possibly the liberation of disk accretion luminosity (e.g., 
Sonnhalter, Preibisch, \& Yorke 1995; Boss \& Yorke 1996; 
Men'shchikov \& Henning 1997; D'Alessio et al. 1999).

The larger flux measurement at 1.3mm is from Reipurth et al. (1993) and 
was observed with a 23'' beam.  
The mm data from Stapelfeldt \& Moneti (1999) are from high spatial 
resolution mapping of the HH30~IRS disk using the Owens Valley Radio 
Observatory ({\it OVRO}).  
As such the {\it OVRO} data probes emission 
from the disk as opposed to the unresolved data of Reipurth et al. (1993).  
The larger flux 
measurement of Reipurth et al. indicates that there is material beyond the 
disk observed by {\it HST} and {\it OVRO}, but the large beam size does not 
spatially resolve this emission or determine whether all of it is really 
associated with HH30~IRS (see also the 1.3mm map of Motte \& Andre 2001, 
their Fig.~1h).  
In our modeling, we are specifically interested in the grain properties 
within the resolved circumstellar disk and therefore our models fit the 
mm data of Stapelfeldt \& Moneti (1999).  The {\it HST} images are fit with 
$R_{\rm disk}\sim 200$AU (Cotera et al. 2001) which is comparable to the 
extent of the mm data we are modeling (Stapelfeldt \& Padgett 2001) and 
we adopt this outer radius (see \S~3.2).  Both 1.3mm points 
are shown in Fig.~1, but in subsequent figures 
only the {\it OVRO} data are shown.

The foreground 
extinction to HH30~IRS is not well determined as the central source 
is not directly visible at optical and near-IR wavelengths.  Cotera et al. 
estimated $A_V < 4$ and Fig.~1 shows dereddened SEDs for 
$1\le A_V \le 4$ using the dereddening formulae of Cardelli, Clayton, 
\& Mathis (1988) with $R_V=3.1$.  Comparing the dereddened SED to 
models of edge-on disks suggests $A_V<2$.

\section{Models}

We now outline the radiation transfer technique, disk geometry, 
luminosity sources, and circumstellar dust properties that are required 
in our modeling.

\subsection{Radiative Equilibrium Calculation}

The radiation transfer calculations use the Monte Carlo 
radiative equilibrium and temperature correction technique of 
Bjorkman \& Wood (2001) which we have adapted to simulate 
T~Tauri disks (Bjorkman, Wood, \& Whitney 2001, in preparation).  
The ouput of 
our code is the disk temperature structure (due to heating by stellar 
photons and accretion luminosity) and the emergent SED at a range of 
viewing angles.  A calculation of the 
hydrostatic disk structure (e.g., D'Alessio et al. 1999) 
can be included in the Monte Carlo technique, but this would require 
an iterative scheme.  At present we have not implemented such a scheme and 
instead perform the radiative equilibrium calculation within a fixed 
disk geometry as we now describe.

\subsection{Disk Structure}

We calculate model SEDs by performing the radiation transfer 
and radiative equilibrium calculation within the disk structure derived 
from scattered light modeling of the {\it HST} images of HH30~IRS.  
The {\it HST} images have been modeled by Burrows et al. (1996), 
Wood et al. (1998), and Cotera et al. (2001) with a flared disk density 
structure,
\begin{equation}
\rho=\rho_0 \left ({R_\star\over{\varpi}}\right )^{\alpha}
\exp{ -{1\over 2} [z/h(\varpi )]^2  }  
\; ,
\end{equation}
where $\varpi$ is the radial coordinate in the disk midplane, 
$\rho_0$ is the density at the stellar surface and the 
scale height increases with radius,
$h=h_0\left ( {\varpi /{R_\star}} \right )^\beta$.  
For our SED modeling we adopt: 
$\beta = 1.25$, $\alpha=2.25$, $h_0=0.017R_\star$, giving 
$h[100{\rm AU}] = 17$AU.  
The $\beta$ and $\alpha$ values are consistent with detailed 
hydrostatic structure calculations (D'Alessio et al. 1999).  
The inner radius, $R_0 = 6R_\star$, is set to be the dust destruction 
radius for $T_\star=3500$K and dust sublimation temperature $T_d=1600$K. 
The outer disk radius is fixed at $R_{\rm disk}=200$AU to match the spatial 
extent of the {\it HST} images and the mm observations of Stapelfeldt 
\& Padgett (2001).  Therefore our SED modeling and derived disk mass and 
dust properties apply to the material within $R_{\rm disk}=200$AU which 
forms the observed SED.  More dust likely exists beyond 200AU 
(Reipurth et al. 1993; Motte \& Andre 2001), but it is 
not detected in the high spatial resolution {\it HST} or mm images and 
therefore does not contribute to the observed SED we are modeling.

To fit the NICMOS F160W image, Cotera et al. (2001) used a slightly 
different geometry ($\beta$ and $\alpha$) from what we described above and 
derived $M_{\rm disk}=6.7\times 10^{-4}M_\odot$.  This scattered light 
modeling probes the mass of small 
particles that dominate the opacity at short wavelengths.   As such, 
modeling of optical and near-IR images is insensitive to large particles, 
rocks, and planetesimals, and therefore this mass is a lower limit.

\subsection{Energy Sources}

The energy input to the disk is from stellar photons and accretion 
luminosity liberated in the disk.  Initially we consider a passive disk 
heated by a uniformly bright star.  \S~4.3 places limits on 
the luminosity arising from viscous accretion.  
As discussed in the previous section, the disk structure is fixed 
for our radiation transfer simulations.  We then 
use $\alpha$ disk theory (Shakura \& Sunyaev 1973) 
to determine the accretion rate for a given disk 
mass and structure.  The accretion rate and viscosity 
parameter, $\dot M$ and $\alpha_{\rm disk}$, are related to the disk 
parameters by
\begin{equation}
\dot M = \sqrt{18 \pi^3}\,\alpha_{\rm disk}\, V_c\, \rho_0\, 
h_0^3/R_\star \; ,
\end{equation}
where the critical velocity $V_c=\sqrt{GM_\star/R_\star}$.  The flux due to 
viscous disk accretion, $GM_\star\dot M/2R_\star$, 
is generated throughout the disk midplane region according to 
\begin{equation}
{{dE}\over{dA\, dt}}={{3GM_\star\dot M}\over{4\pi \varpi^3}}
\left [ 1-\sqrt{{R_\star}\over{\varpi}}\right ]\; .
\end{equation}
We adopt $M_\star = 0.5M_\odot$ and vary $\alpha_{\rm disk}$ to place 
limits on the accretion luminosity for a given $M_{\rm disk}$.

The variability of HH30's scattered light nebula (Burrows et al. 1996; 
Stapelfeldt et al. 1998; Cotera et al. 2001) and ground based $VRI$ 
photometry (Wood et al. 2000) 
indicates non-uniform illumination of the disk due to either 
stellar hotspots or shadowing due to large inhomogeneities in the inner 
disk.  However, D'Alessio et al. (1998) showed that hot accretion rings do 
not significantly alter the SED, so we assume a uniformly bright star 
and do not include disk warping in our models.

\subsection{Circumstellar Dust Properties}

Burrows et al. fitted their HH30~IRS {\it HST} images 
using a dust scattering phase function that was more forward throwing than 
ISM dust grains and the near-IR modeling by Cotera et al. derived a 
shallower wavelength dependent opacity than that of ISM grains.  
Both these results are indicative of circumstellar dust grains 
that are larger (in a statistically significant sense, e.g., cross-section
weighted mean size) than typical ISM grains.  Grain growth within 
protoplanetary disks has been inferred from mm observations and the slope 
of long wavelength SEDs --- mm fluxes are larger and the SED slopes are 
shallower than predicted from ISM grains (e.g., Beckwith et al. 1990; 
Beckwith \& Sargent 1991).  
In order to examine the possibility of grain growth within the HH30 disk
as well as the potential characteristics thereof, 
we construct models with several different dust size distributions: 
a standard ISM mixture and several distributions with larger grains
including both simple power-laws and power-laws with an exponential
cutoff.

For the ISM dust, we adopt the size (mass) distribution of Kim, Martin,
\& Hendry (1994; KMH) for the canonical diffuse interstellar sightline
(i.e, $R_V$=3.1).  In addition, we also reproduce the model of
Cotera et al. (2001).  The interested reader is referred to these two works
for specific details of the models.

For our ``larger grain'' dust models, we 
combine a ``power-law with exponential cutoff'' (e.g., KMH) for
components of amorphous carbon and astronomical silicates with solar
abundance constraints for carbon and silicon (C/H $\sim 3.5 \times 10^{-4}$
and Si/H $\sim 3.6 \times 10^{-5}$, respectively; Anders \& Grevesse 1989;
Grevesse \& Noel 1993).  The primary motivation for introducing an additional
complexity onto a simple power-law (one with a single exponent, e.g.
D'Alessio et al. 2001) is that it is not
expected to apply over such a large range of grain sizes.  This has been
indicated in analyses of both ISM
and circumstellar dust grains (e.g., KMH; Jones, Tielens, \& Hollenbach 1996;
Witt, Smith, \& Dwek 2001; Suttner \& Yorke 2001).  In addition, rather
than joining two separate power-laws (i.e.,
Witt, Smith, \& Dwek 2001), we adopt a distribution that provides a
smooth transition between two different regimes and that can reproduce
the general behavior of a large range of interstellar sight lines (e.g., KMH).
The functional form of this size distribution is :
\begin{equation}
n(a) \,\, da  =  C_i \, a^{-p}\,  \exp{(-[a/a_c]^q)} \,\ da\; .
\end{equation}
The parameters $p$, $a_c$, $q$, which control the distribution shape, 
are adjusted to fit the wavelength dependence of the observations. Of
course, a simple power-law corresponds to the case of $q=\infty$.
$C_i$ is set by
requiring the grains to completely deplete a solar abundance of carbon
and silicon.  The subscript ``$i$'' refers to the different components of 
the dust model: amorphous carbon and silicate (see Table~1).  
This function is essentially
the ``modified gamma distribution'' found in atmospheric
and planetary science literature.  The end-points of the size
integration, $a_{\rm min}$ and $a_{\rm max}$, are also parameters.
Not surprisingly, models with finite values of $q$ are insensitive
to the exact value of $a_{\rm max}$ as long as it is $ >> a_c$.

The opacity of the 
dust grains is modelled under the assumption of two separate
populations of homogeneous, spherical particles.  In general, we use
published dielectric functions for astronomical silicate (Weingartner \&
Draine 2001) and amorphous carbon (BE type, Zubko et al. 1996).  However, our
desired wavelength range extends longward of both sets of dielectric
functions and slightly shortward of the amorphous carbon values (i.e., 
100-500 \AA\ not in published set).  We use a subtractive Kramers-Kronig
analyis (cf. Ahrenkiel 1971; routine kindly supplied by K. Snook),
specifying the imaginary index of refraction in the regions of interest.
This approach also requires specification of the real index of refraction 
at some reference point; we use a value of the real
refractive index at a wavelength adjacent to the missing range from
the published data.  For the long wavelength extension, we use the
approximation that imaginary index is proportional to $\lambda^{-N}$
(e.g., Pollack et al. 1994): $N=-1$ for the silicates and $N=1/6$ for
the amorphous carbon.  The latter choice departs from the recommendation
of Pollack et al. for metals ($N=1$), but is a much better match for the
slope of the published data in the far-IR region.  Fortunately, our
model fits are fairly insensitive to the exact choice of $N$.
For the short wavelength extension, we use the imaginary index of the
BE1 of Rouleau \& Martin (1991) scaled to match smoothly with that of 
Zubko et al.

The scattering properties are calculated using a combination of
Mie series solution and geometrical optics algorithms.  For
the former, we employ the code of Wiscombe (1996) while the
latter regime is handled with the code of Macke and
collaborators ({\it rt-ellipsoid}, i.e., Macke \& Mishchenko 1996). 
The transition point between the two algorithms is made at a size
parameter of 20,000.  Though this choice is somewhat arbitrary,
at this point the forward scattering properties as calculated by each
algorithm differ at a level below 1\% (the phase matrix elements are
in agreement at the 5-10\% level).  This is also true at a
size parameter of 10,000, but we chose the larger value in order
to allow the phase functions in the visible and NIR to be calculated
entirely by the Mie algorithms; this avoids introducing the small-scale
($\sim$ 10\% outside the central diffraction peak)
high-frequency oscillations produced by the geometrical optics algorithm.

Figures~2 and 3 show the spectral shape of the opacity, albedo, and cosine
asymmetry parameter (i.e., $g$) for the various dust models we consider.  
The size distribution parameters are listed in Table~1.
Although we use Mie theory for calculating the scattering properties, in 
the radiative equilibrium calculation we adopt a simple Heyney-Greenstein 
scattering phase function characterized by the single parameter, $g$ 
(Heyney \& Greenstein 1941).

\section{Model SEDs}

A tentative spectral type of M0 for the illuminating source of HH30~IRS 
was assigned by Kenyon et al. (1998).  However, because the central star 
is totally obscured by the disk at optical and near-IR wavelengths, the 
spectral type is very uncertain.  Scattered light models indicate 
$L_\star \approx 0.2L_\odot$, so we adopt $T_\star = 3500$K (Kurucz model 
atmosphere, Kurucz 1994) and $R_\star = 1.2R_\odot$.  The distance to 
HH30~IRS is taken to be 140pc.  
The following sections show the resulting SEDs 
for different dust models, starting with a purely passive reprocessing disk.  
For each dust model we adopt the disk size and structure described in 
\S3.2 and determine the disk mass from fitting the NICMOS F160W image.  
We then calculate the model SED for comparison with the data presented in 
figure~1.  Figure~4 shows the NICMOS F160W image and scattered light model 
fits for the five dust types described in \S3.4 and summarized in Table 2.  
The $A_V$ values in this table are calculated differently than those 
presented by Cotera et al. (2001).  We have simply integrated through the 
disk along a viewing angle of $i=84^\circ$, whereas Cotera et al. measured 
how much light escapes in the Monte Carlo simulation to get an effective 
$A_V$.  

The width of the dust lane in the scattered light images is sensitive not 
only to the disk mass (and therefore $\tau_{eq}$and $A_H$), 
but also to the scattering phase function.   If $g$ were the same for 
all models then $A_H$ would be the same also.  However, as the phase 
function becomes more forward throwing for the 
larger dust grain models (i.e., $g$ increases), a larger optical 
depth is required to obtain the same width of the dust lane. 

\subsection{Reprocessing Disk SEDs for $i=84^\circ$}

Figures~5 and 6 show the HH30~IRS data and model SEDs 
(for $i=84^\circ$ derived 
from scattered light models) for the dust types 
described in \S3.4 and summarized in Table~1.  
The models show the characteristic double 
peaked SEDs of highly inclined disks.  The disk mass has been chosen so 
that each model matches the NICMOS F160W image (Cotera et al. 2001).  
The different dust models are 
distinguished from one another at longer wavelengths.  Although the ISM 
grains and the Cotera et al. grains produce similar mm fluxes, they differ 
in the $1-10\mu{\rm m}$ range where the ISM grains have a steeper 
wavelength dependent 
opacity that yields too narrow a dust lane in NICMOS scattered light 
images.  The mm fluxes are best reproduced with larger grain size 
distributions.  Our pure reprocessing disk models yield a rather low 
flux in the $1-10\mu {\rm m}$ range and \S4.3 shows models including 
accretion luminosity that provide a better fit to this spectral regime.

The different models, disk mass estimates, midplane optical depths 
($\tau_{eq}$), and extinction to the star for $i=84^\circ$ are summarized in 
Tables~1 and 2.  The simple power-law size distributions 
can have very large mm optical depths.  
While dust size distributions with small $a_{\rm max}$ can be used to 
fit individual scattered light images, the mm fluxes are best fit with 
the large grain models.  These models require very different 
masses and our SED models show that they may be distinguished in the 
mid to far-IR measured by {\it SIRTF} and in the sub-mm measured by 
{\it SMA} and {\it ALMA}.  That is to say, although a simple power-law
is not physically plausible in objects like HH30~IRS, the issue
of whether it adequately reproduces the SED can be determined with
sufficient observation coverage.

\subsection{Reprocessing Disk SEDs for All Inclinations}

Scattered light modeling of HH30~IRS indicates that we are seeing a 
Classical T~Tauri star viewed almost edge-on.  Whitney \& Hartmann (1992) 
presented scattered light disk models for a range of viewing angles 
showing that edge-on disks, where the central star is obscured, will be most 
easily detectable.  Our edge-on SED models agree well with the current 
HH30~IRS data and Fig.~7 shows model SEDs at a range of 
viewing angles.  As expected, for viewing angles where the central source is 
unobscured, our HH30~IRS models do indeed resemble the SED of a Classical 
T~Tauri star.  
The silicate features at $9\mu{\rm m}$ and $18\mu{\rm m}$ 
are most prominent in the ISM dust model.  As the grain size increases 
and the opacity is dominated by larger grains, the relative strengths of 
the silicate features decrease (Fig.~2).  This is reflected in the SEDs 
with the $a_c=50\mu{\rm m}$ size distribution showing a very smooth 
SED across the silicate features.  However, at present our models do 
not allow for different dust compositions at different locations within 
the disk as required for a detailed investigation of the disk chemistry 
and mineralogy through analysis of crystalline features 
(e.g., Chiang et al. 2001).

\subsection{Effects of Accretion Luminosity}

HH30~IRS is the source of a powerful jet (Mundt \& Fried 1983; 
Mundt et al. 1990; Graham \& Heyer 1990; Lopez et al. 1995), 
presumably driven by accretion 
luminosity.  The low disk mass for HH30~IRS derived using ISM dust models 
is unlikely to 
support a large accretion rate in the context of $\alpha$ disk theory.  
Therefore the larger disk masses we derive are more appealing as they allow 
for significant accretion rates.   Figure~8 shows models which include 
disk heating from viscous accretion for Model~1, which has 
$M_{\rm disk}=1.5\times 10^{-3}M_\odot$.  

We find $L_{\rm acc} = 0.2L_\star$ provides a good match to the 
$1-10\mu {\rm m}$ fluxes presented by Brandner et al. (2000).  Assuming 
$M_\star=0.5M_\odot$ then $\dot M = 4\times 10^{-9}M_\odot {\rm yr}^{-1}$ 
and $\alpha_{\rm disk}=0.02$.  A larger accretion luminosity produces too 
much $1 -10\mu{\rm m}$ emission compared to the observations.  
Figure~8 also shows models in which the accretion luminosity is lower 
($L_{\rm acc}/L_\star = 5$\%, 2\%).  Pole-on viewing yields 
typical near-IR excesses arising from accretion disks.  

The large uncertainties in the 
$1-10\mu {\rm m}$ fluxes in addition to the unknown $M_\star$ and 
$\alpha_{\rm disk}$ 
preclude an accurate determination of the accretion rate.  However, 
the accretion rate we derive is comparable to the total 
outflow rate in the jets, 
$\dot M_{\rm jet}=3.5\times 10^{-9}M_\odot {\rm yr}^{-1}$ 
estimated by Bacciotti, Eisloffel, \& Ray (1999).  
The combination of SED models of more accurate {\it SIRTF} fluxes and 
spectral line accretion diagnostics (e.g., Gullbring et al. 1998) are 
required for a more accurate determination of the accretion rate.  

\subsection{Components of the SED}

A feature of Monte Carlo radiation transfer is that we may track the 
history of each individual photon packet within the simulation and 
determine the relative contributions to the total SED of direct, 
scattered, reprocessed, and accretion luminosity.  Figure~9 shows 
model SEDs for dust Model~1 at inclinations 
$i=0^\circ$, $i=60^\circ$, $i=71^\circ$, and 
$i=84^\circ$ along with their component 
parts.  The scattered light comprises stellar photons that have been 
(multiply) scattered; thermal photons are stellar photons that have been 
absorbed, reprocessed, and possibly scattered by the disk; 
accretion photons are those that initiated as accretion luminosity, but 
may have subsequently been scattered or absorbed and re-radiated.

For pole-on viewing, the SED is dominated by direct starlight 
at optical wavelengths and thermal (i.e., reprocessed) disk emission 
at long wavelengths.  Accretion luminosity contributes significantly 
at near and mid-IR wavelengths, while scattered starlight contributes 
a few percent of the SED in the optical and near-IR and much less at 
longer wavelengths.  Scattering will always yield an 
excess above the photospheric level (see also Whitney \& Hartmann 1992).  
For edge-on viewing, the situation is different: 
the star is totally obscured by the disk and scattered starlight dominates 
the SED in the optical and near-IR.

\section{Summary}

We have presented Monte Carlo radiation transfer calculations for the SED 
of HH30~IRS.  Our model comprises a central stellar source surrounded
by a circumstellar disk whose structure was derived from modeling 
{\it HST} images.  The dust opacity in the disk has a
shallower wavelength dependence than typical ISM dust and
the average grain size is significantly larger.  The model that 
matches the observed SED has a dust size distribution extending to larger 
grain sizes than ISM dust; specifically $a_{\rm max}=1$mm with an 
exponential cutoff scale-length of $a_c=50\mu{\rm m}$.  The opacity at 
long wavelengths is roughly 
$\kappa\sim\lambda^{-1}$, in agreement with that derived from sub-mm 
continuum SEDs in many classical T~Tauri stars (Beckwith et al. 1990). 
While size distributions with most of the mass in smaller grains can be
constructed to fit optical and near-IR 
images, these grain models produce too little mm emission. 
The disk mass we derive using the model above is
$M_{\rm disk}=1.5\times 10^{-3}M_\odot$, considerably 
larger than the $M_{\rm disk}=2\times 10^{-4}M_\odot$ found by Wood et al. 
(1998) using ISM grains.  

The derived disk mass is a lower limit, since the SED 
is not very sensitive to large particles, such as rocks 
and planetesimals.  Estimations of disk mass from mm observations are 
uncertain due to the poorly known opacity at mm wavelengths.  For 
HH30's disk, Reipurth et al. (1993) estimated $M_{\rm disk}=0.02M_\odot$ 
if $\kappa_{\rm 1.3mm}=0.003{\rm cm}^2/{\rm g}$ or 
$M_{\rm disk}=0.003M_\odot$ if $\kappa_{\rm 1.3mm}=0.02{\rm cm}^2/{\rm g}$.  
The opacity for our grain size distribution is 
$\kappa_{\rm 1.3mm}=0.08{\rm cm}^2/{\rm g}$, again demonstrating 
the sensitivity of the disk mass estimate to the adopted mm opacity.
Observations with {\it SIRTF} and at sub-mm wavelengths will give a
more complete picture of the SED of HH30~IRS and other edge-on disks, 
allowing us to further constrain the
dust size distribution and ascertain the extent of the grain growth
within the disk.  In particular, the SED in the $20-800\mu{\rm m}$ range
will discriminate between the dust model with exponential
cut-off (corresponding to $M_{\rm disk}=1.5\times 10^{-3}M_\odot$) and
the single power-law size distributions which require much more massive
disks ($M_{\rm disk}\sim 0.5M_\odot$).

Grain growth in protoplanetary disks has been inferred from observations:
e.g., shallower slopes of the long wavelength SEDs and larger mm fluxes 
compared with those expected for ISM dust parameters (Beckwith et al. 1990; 
Beckwith \& Sargent 1991);
the observed wavelength dependence of dust lane widths in near-IR images of
HH30~IRS (Cotera et al. 2001).  Our SED modeling of HH30~IRS provides further 
evidence for grain growth and suggests larger grain 
sizes than those derived from scattered light models only.  
Grain growth within protoplanetary disks is expected on timescales of 
$10^5$ years (Beckwith, Henning, \& Nakagawa 2000 and references therein).  
The simulations of Suttner \& Yorke (2001)
find that compact grains can grow to $a=50 \mu{\rm m}$ within $\sim 10^4$
years.  In their model, however, the largest grains are found primarily in
the disk midplane regions and as such will not have a big influence on the
infrared SED, which is dominated by the upper layers or disk ``atmosphere.''
Although the current implementation of 
our code adopts the same grain size distribution throughout the disk, 
our SED modeling is consistent with grain growth to larger than
$50\mu{\rm m}$, even in the upper layers of the disk atmosphere that 
dominate the infrared SED.

The mass that we derive for HH30's disk is more appealing than the low 
disk mass derived from ISM dust models which are unlikely to support 
large accretion rates.  More massive disks support larger 
accretion rates and for HH30~IRS we infer an accretion luminosity of 
$L_{\rm acc}\approx 0.2L_\star$.  Our derived accretion rate, 
$\dot M = 4\times 10^{-9}M_\odot {\rm yr}^{-1}$, is uncertain as it 
depends on the unknown $M_\star$ and $\alpha_{\rm disk}$ 
viscosity parameter and the accurate determination of $\dot M$ 
requires spectral line observations in addition to SED modeling.

\acknowledgements

We acknowledge financial support from NASA's Long Term Space Astrophysics 
Research Program, NAG5~6039 (KW), NAG5~8412 (BW), NAG5~7993 (MW), 
NAG5~3248 (JEB);
the National Science Foundation, AST~9909966 (BW and KW), AST~9819928 (JEB), 
and a PPARC Advanced Fellowship (KW).  
We thank
Debbie Padgett and Karl Stapelfeldt for providing their {\it ISO} and
{\it OVRO} SED data.

\newpage

\begin{deluxetable}{lccccccc}
\tablenum{1}
\tablewidth{0pt}
\tablecaption{Dust Model Parameters}
\tablehead{
\colhead{Dust Model} & \colhead{Component} & \colhead{$C_i$} &
\colhead{$p$} & \colhead{$q$} 
& \colhead{$a_c (\mu{\rm m})$} & \colhead{$a_{\rm max}(\mu{\rm m})$}
}
\startdata
Cotera\dotfill & amc & $5\times 10^{-16}$   & 3.5 & 1.0 & 0.55 & 20 \\
               & sil & $6\times 10^{-16}$   & 3.0 & 1.0 & 0.55 & 20 \\
Model 1\dotfill & amc & $1.32\times 10^{-17}$   & 3.0 & 0.6 & 50 & 1000  \\
               & sil & $1.05\times 10^{-17}$   & 3.0 & 0.6 & 50 & 1000   \\
Model 2\dotfill & amc & $1.32\times 10^{-17}$   & 3.5 &$\infty$  &\nodata& 1000 \\
               & sil & $1.05\times 10^{-17}$     & 3.5 &$\infty$  &\nodata& 1000 \\
Model 3\dotfill & amc & $1.32\times 10^{-17}$   & 3.0 &$\infty$  &\nodata& 1000  \\
               & sil & $1.05\times 10^{-17}$     & 3.0 & $\infty$ &\nodata& 1000  \\
\enddata
\tablenotetext{ }{The size distribution for the KMH model is taken from
their mass distribution for the fit to the canonical ($R_V$=3.1)
average diffuse ISM sight line.}
\end{deluxetable}

\begin{deluxetable}{lccccccc}
\tablenum{2}
\tablewidth{0pt}
\tablecaption{Disk Masses and Extinctions}
\tablehead{
\colhead{Dust Model} & \colhead{$M_{\rm disk}/M_\odot$} &
\colhead{$A_H$} & \colhead{$\tau_{eq}(H)$} 
& \colhead{$g(H)$} & \colhead{$A_V$} & \colhead{$A_{1.3{\rm mm}}$}
}
\startdata
KMH\dotfill     & $3.5\times 10^{-4}$ & 800   & 31000  & 0.28 & 4600 & 0.05\\
Cotera\dotfill  & $3.5\times 10^{-4}$ & 1000  & 39000  & 0.49 & 2800 & 0.05\\
Model 1\dotfill  & $1.5\times 10^{-3}$& 1800  & 65000  & 0.58 & 3300 & 7.9\\
Model 2\dotfill  & $0.015$            & 3700  & 130000 & 0.60 & 6500 & 120\\
Model 3\dotfill  & $0.5$              & 13000 & 440000 & 0.75 & 15000& 4200\\
\enddata
\tablenotetext{}{Optical depths are in the disk midplane, 
extinctions are along $i=84^\circ$.}
\end{deluxetable}

\begin{figure}[t]
\centerline{\plotfiddle{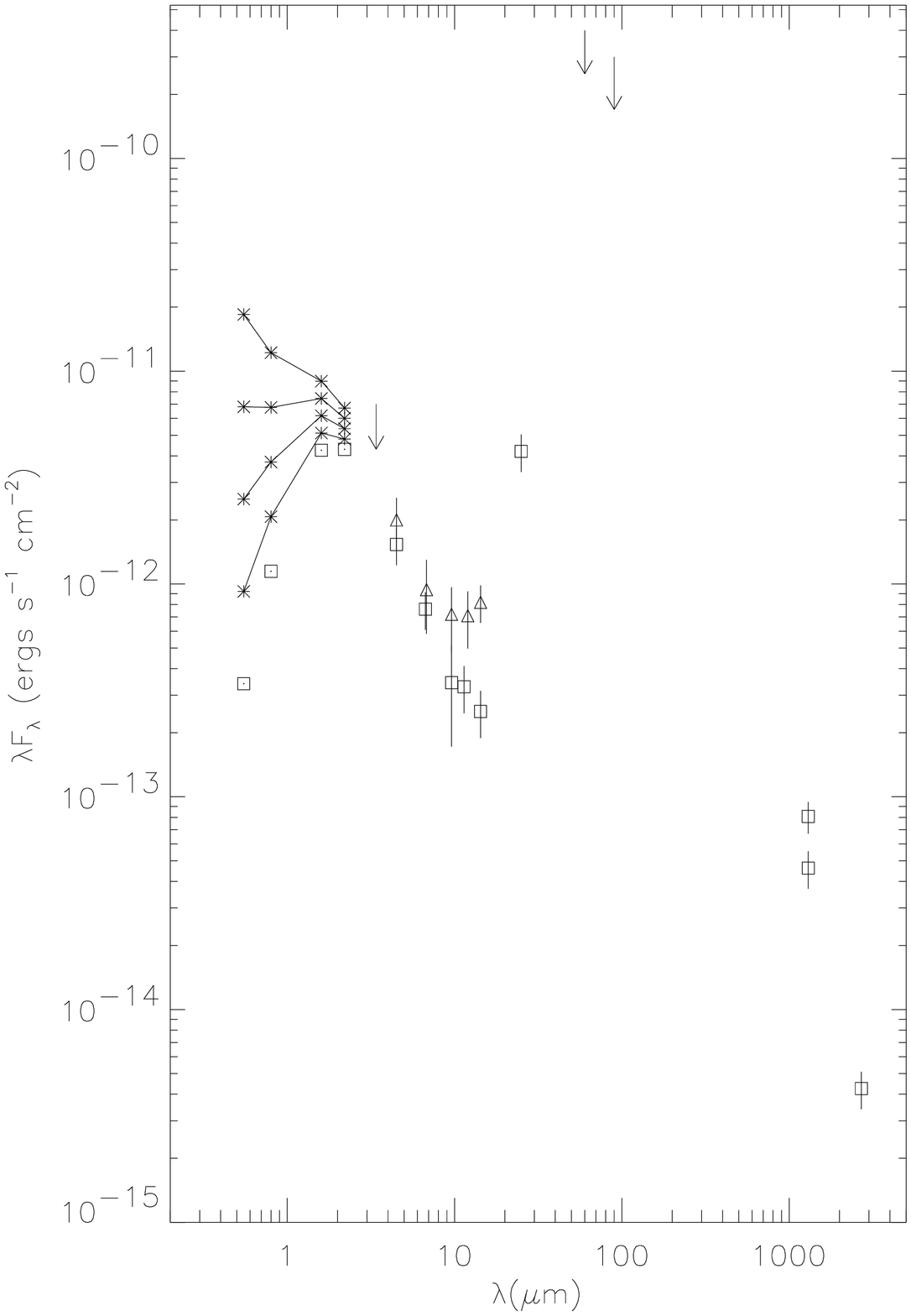}{6in}{0}{65}{65}{-420}{0}}
\caption{Spectral energy distribution of HH30~IRS.  Triangles are the 
{\it ISO} reductions of Brandner et al. (1999).  The {\it ISO} fluxes 
at 60$\mu$m and 90$\mu$m are upper limits.  Also shown are dereddened SEDs 
(stars) assuming $1\le A_V\le 4$, with $A_V=1$ the lowest curve.}
\end{figure}

\begin{figure}[t]
\centerline{\plotfiddle{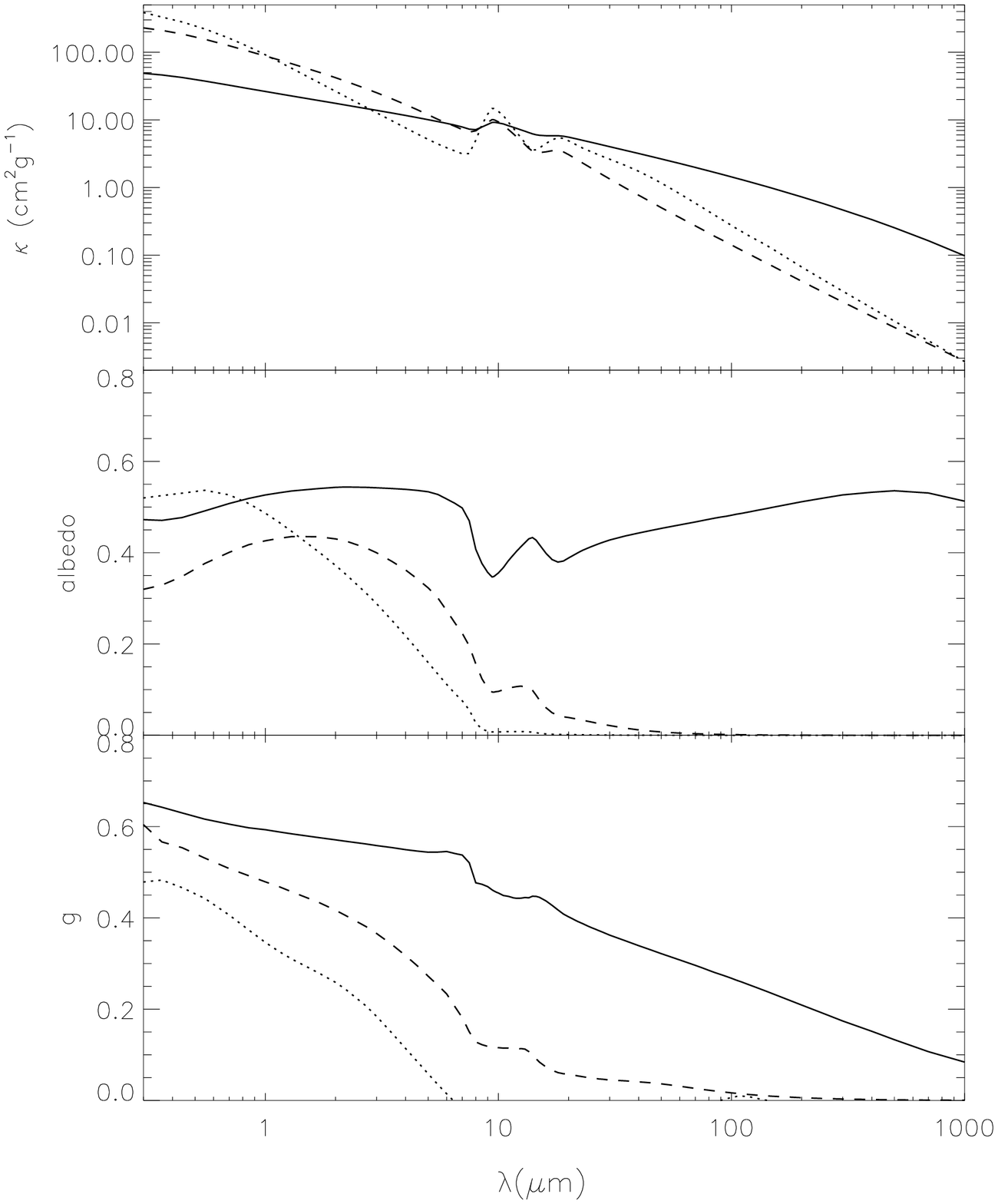}{6in}{0}{65}{65}{-470}{0}}
\caption{Dust parameters for ISM grains (dotted lines), size distribution 
used by Cotera et al. (dashed lines), and the size distribution (Model 1)
we use for fitting the HH30 SED (solid lines).}
\end{figure}

\begin{figure}[t]
\centerline{\plotfiddle{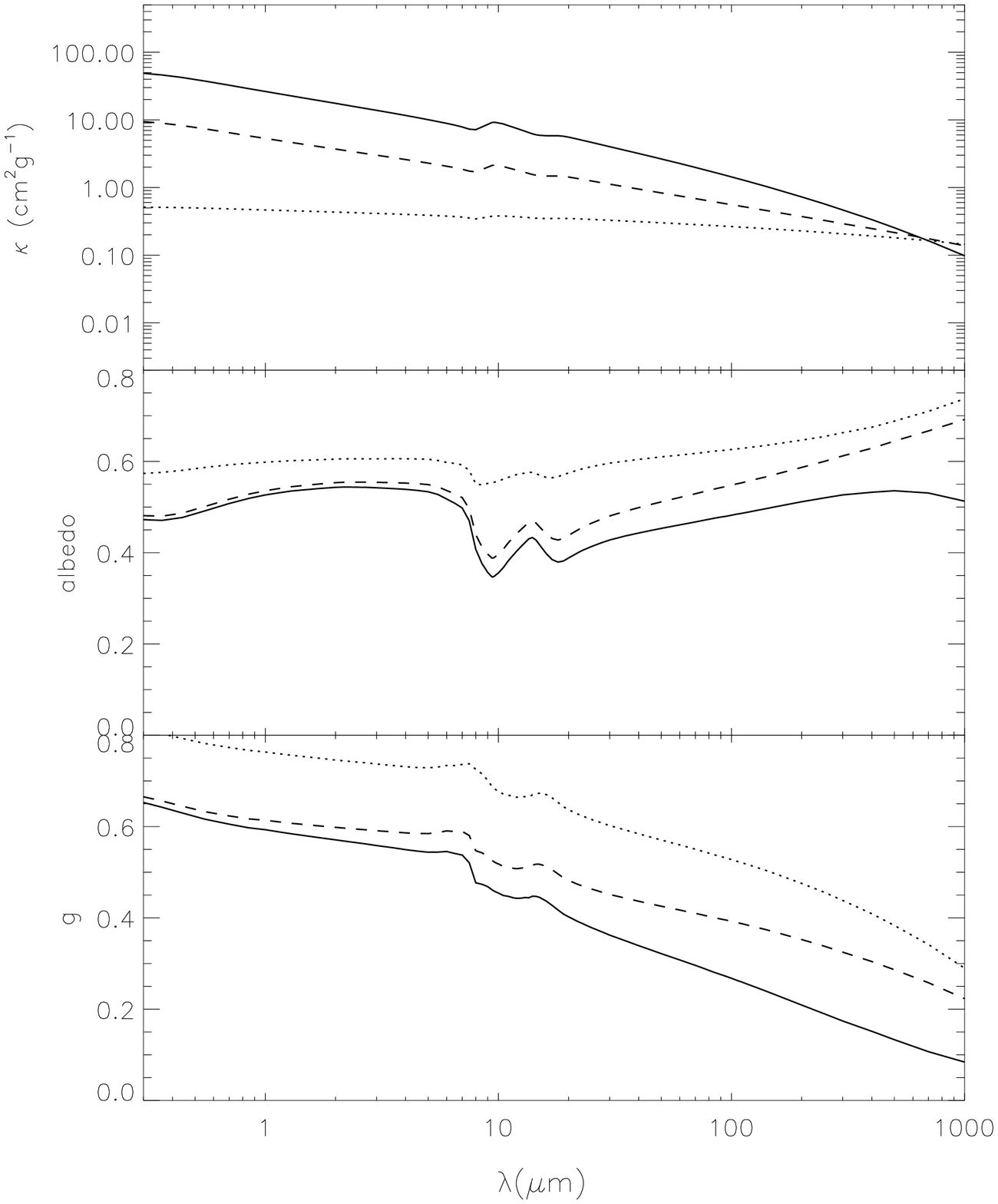}{6in}{0}{65}{65}{-470}{0}}
\caption{Dust parameters for the exponential cutoff dust model (solid line, 
Model 1) 
and simple power-law models, $p=3.5$ (dashed line, Model 2), $p=3.0$
(dotted line, Model 3).}
\end{figure}

\begin{figure}[t]
\centerline{\plotfiddle{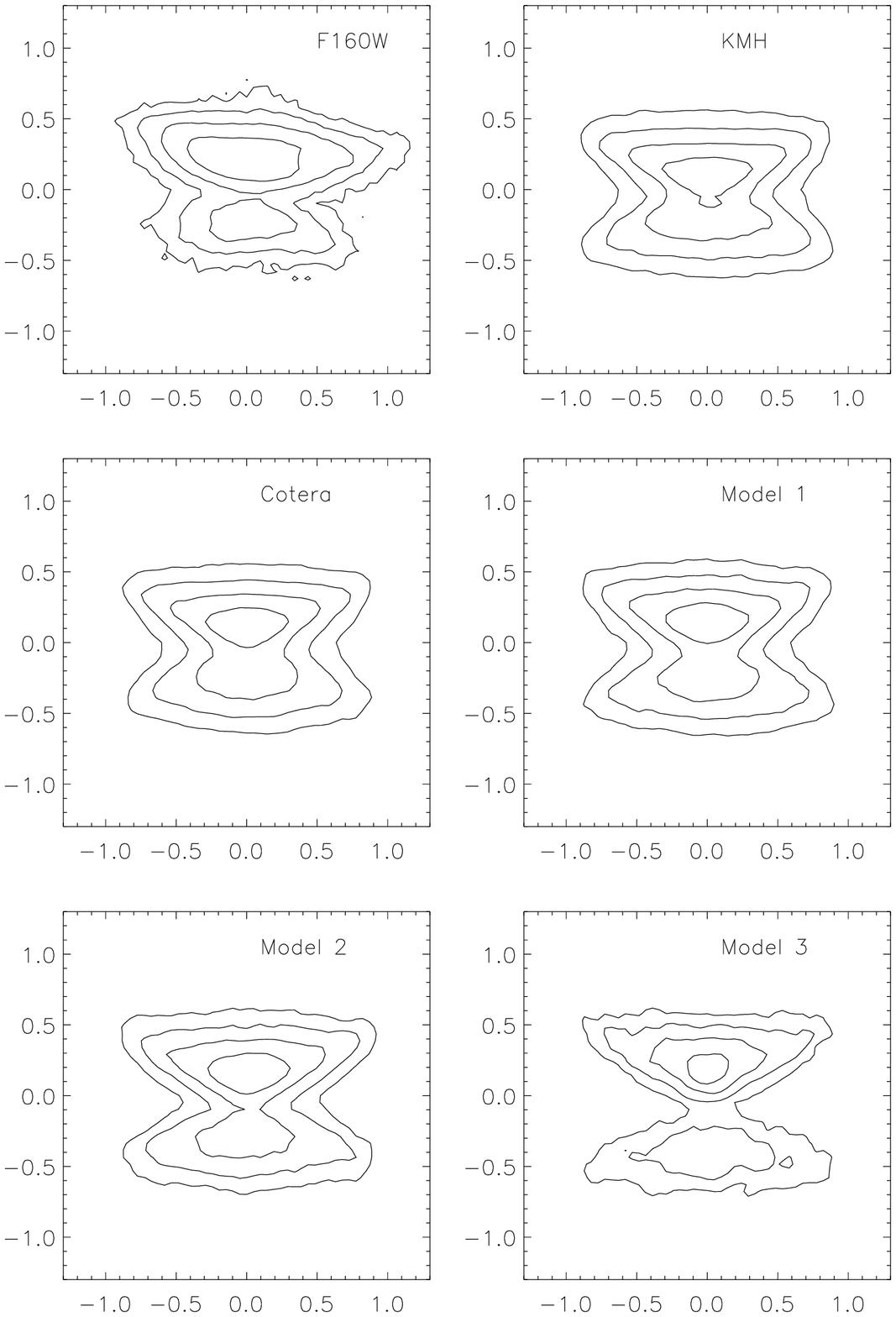}{6in}{0}{65}{65}{-470}{0}}
\caption{NICMOS F160W image of HH30~IRS and scattered light models for the 
different dust size distribution models investigated in this paper.  Axes 
are labeled in arcsecs, where $1{\rm arcsec}=140$AU.}
\end{figure}

\begin{figure}[t]
\centerline{\plotfiddle{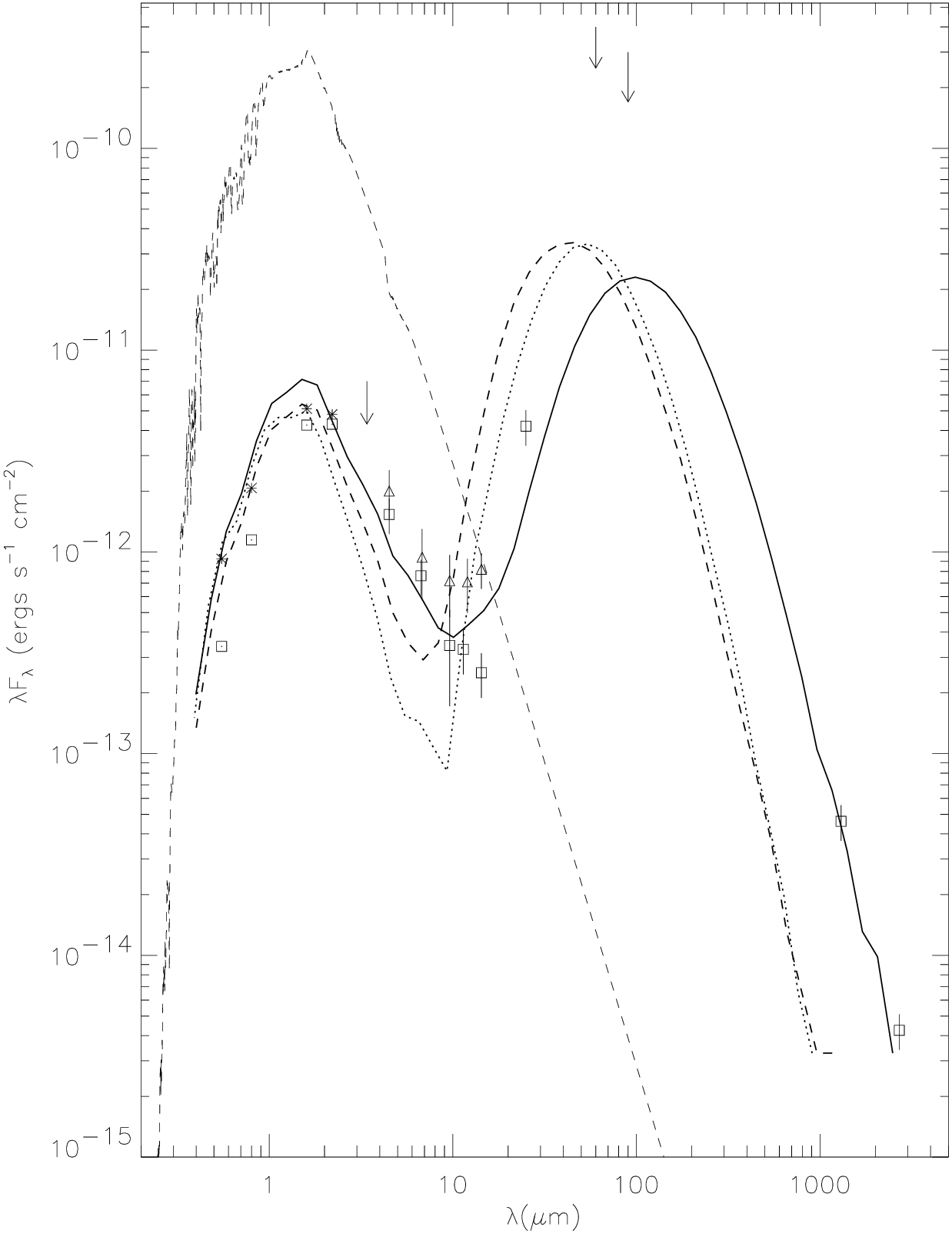}{6in}{0}{65}{65}{-420}{0}}
\caption{Data and model SEDs for passive disks for the three different dust 
models presented in Fig.~2.  Stars show dereddened fluxes assuming 
$A_V=1$.  Model SEDs for $i=84^\circ$ 
are for ISM grains (dotted line), Cotera et al. grains 
(dashed line), exponential cutoff (solid line, Model 1).  The input stellar 
spectrum is shown and the adopted source distance is 140pc.  }
\end{figure}

\begin{figure}[t]
\centerline{\plotfiddle{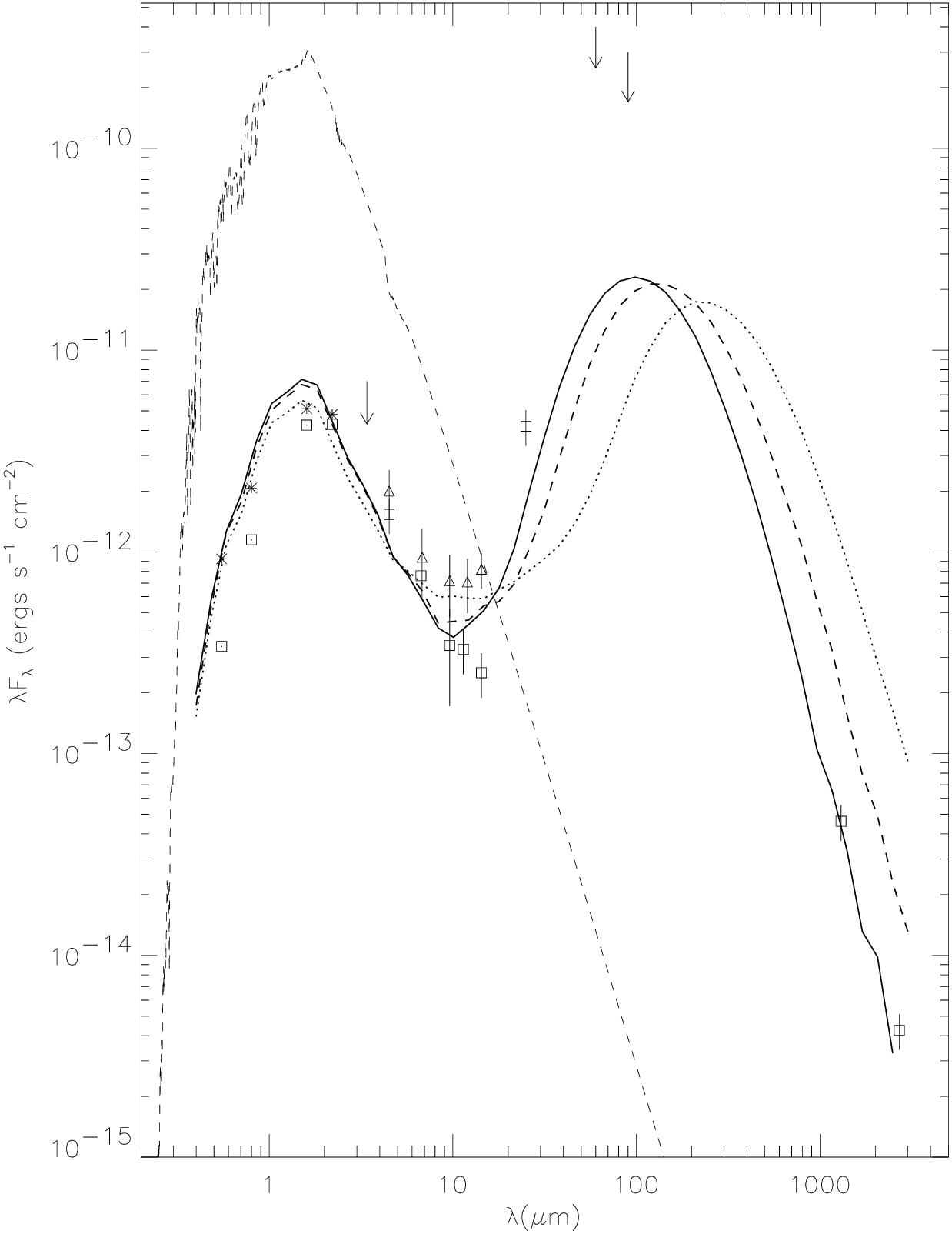}{6in}{0}{65}{65}{-420}{0}}
\caption{Data and model SEDs for the three different dust 
models presented in Fig.~3.  Stars show dereddened fluxes assuming 
$A_V=1$.  Model SEDs are for the exponential cutoff (solid line, Model 1),
and simple power-laws, $p=3.5$ (dashed line, Model 2), $p=3.0$
(dotted line, Model 3).  The input stellar spectrum is also shown. 
The models may be discriminated by the SED in the range
$20\mu{\rm m}\la\lambda\la 1$mm.}
\end{figure}

\begin{figure}[t]
\centerline{\plotfiddle{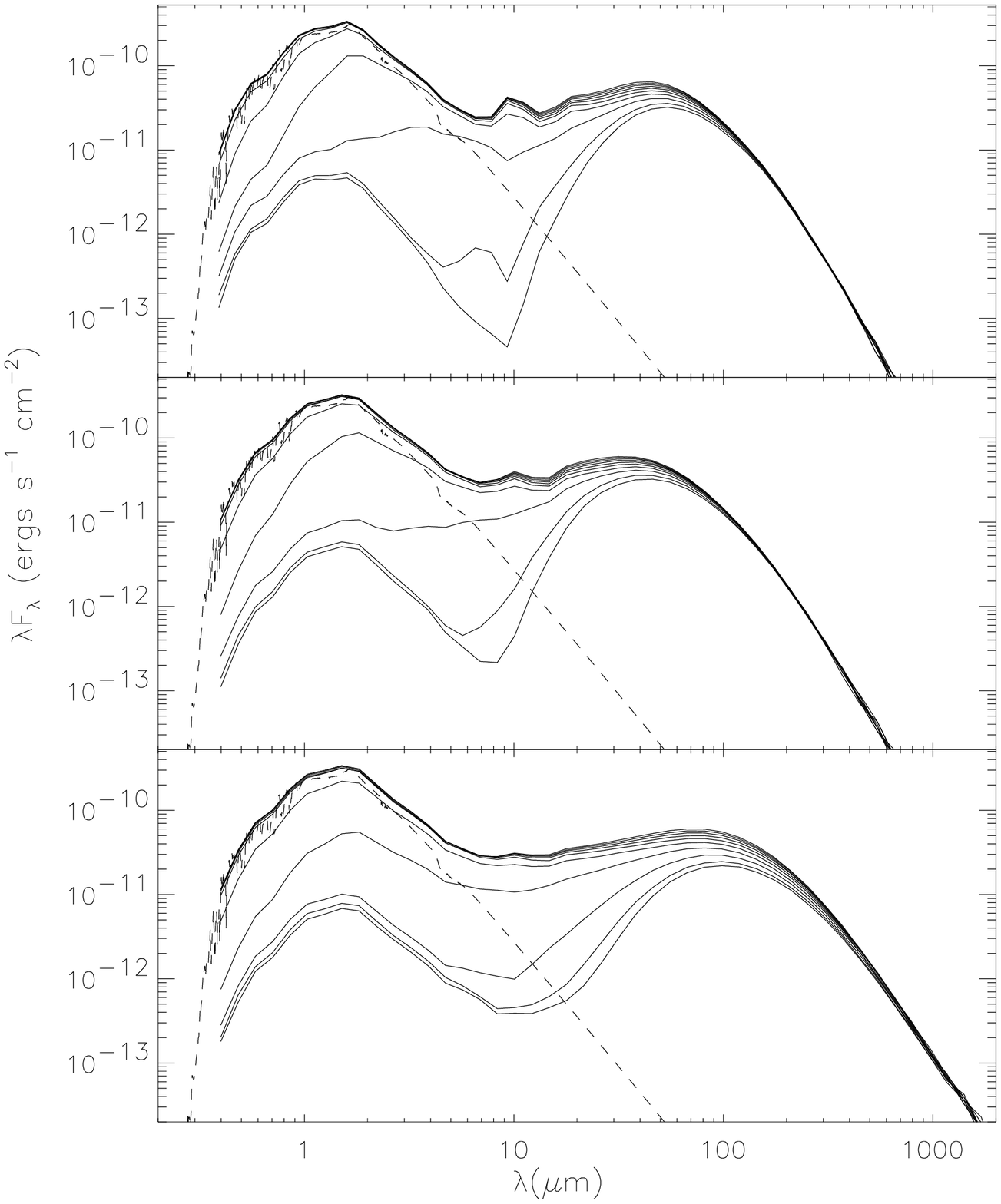}{6in}{0}{65}{65}{-420}{0}}
\caption{Model SEDs at a range of viewing angles 
for KMH grains (upper panel), Cotera et al. (middle panel) and the 
size distribution we use for fitting the HH30~IRS SED (lower panel, 
Model 1).  
Notice the decrease in the strength of the silicate features at $9$ 
and $18\mu{\rm m}$ for the larger grain models.}
\end{figure}

\begin{figure}[t]
\centerline{\plotfiddle{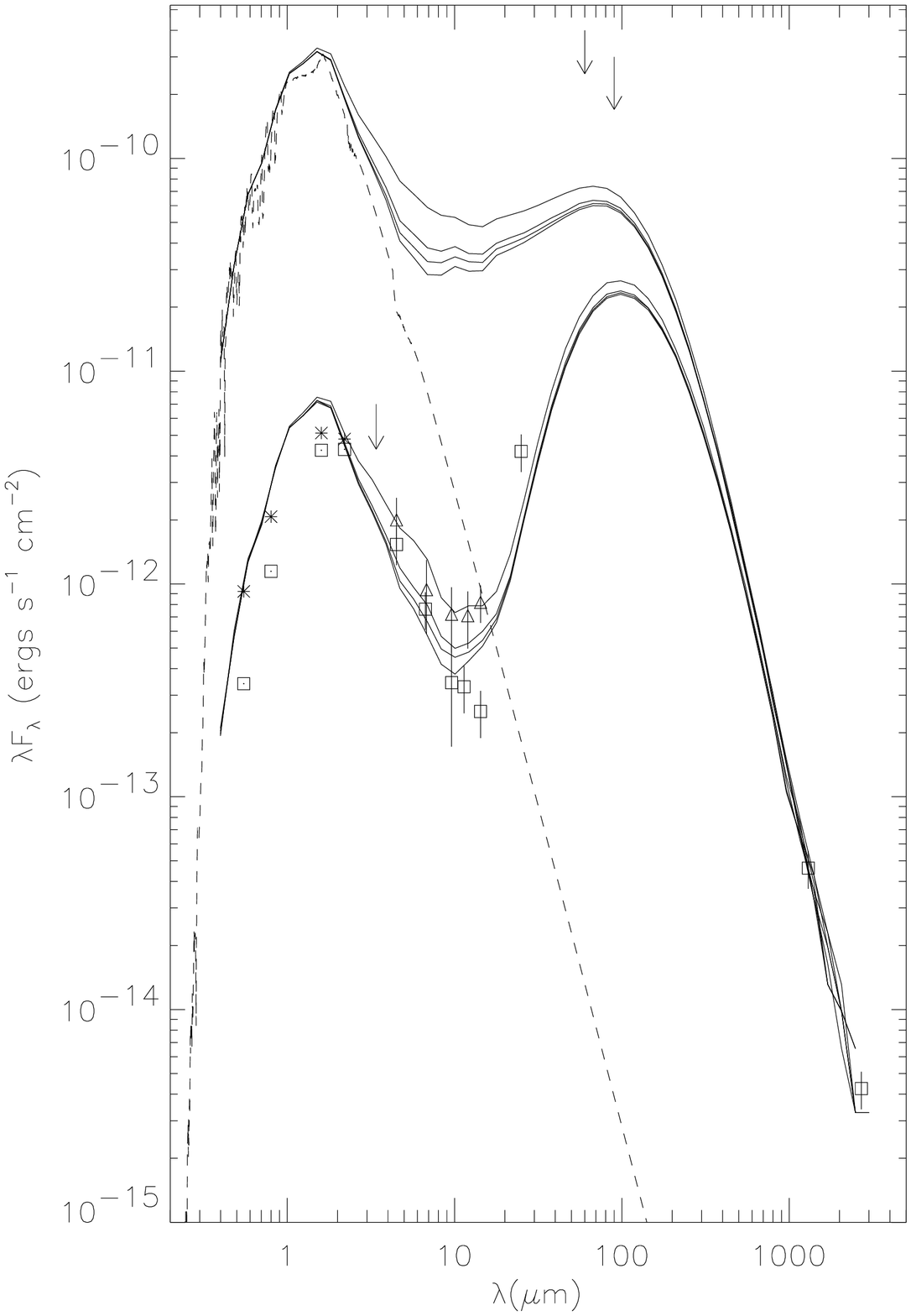}{6in}{0}{65}{65}{-420}{0}}
\caption{Data and model SEDs including accretion luminosity.  
The model SEDs are for $i=84^\circ$ (lower curves) and pole-on models 
(upper curves).  For the different viewing angles going from upper to 
lower curves the accretion luminosity is $L_{\rm acc}/L_\star = 0.2$, 
0.05, 0.05, and 0.0.  }
\end{figure}

\begin{figure}[t]
\centerline{\plotfiddle{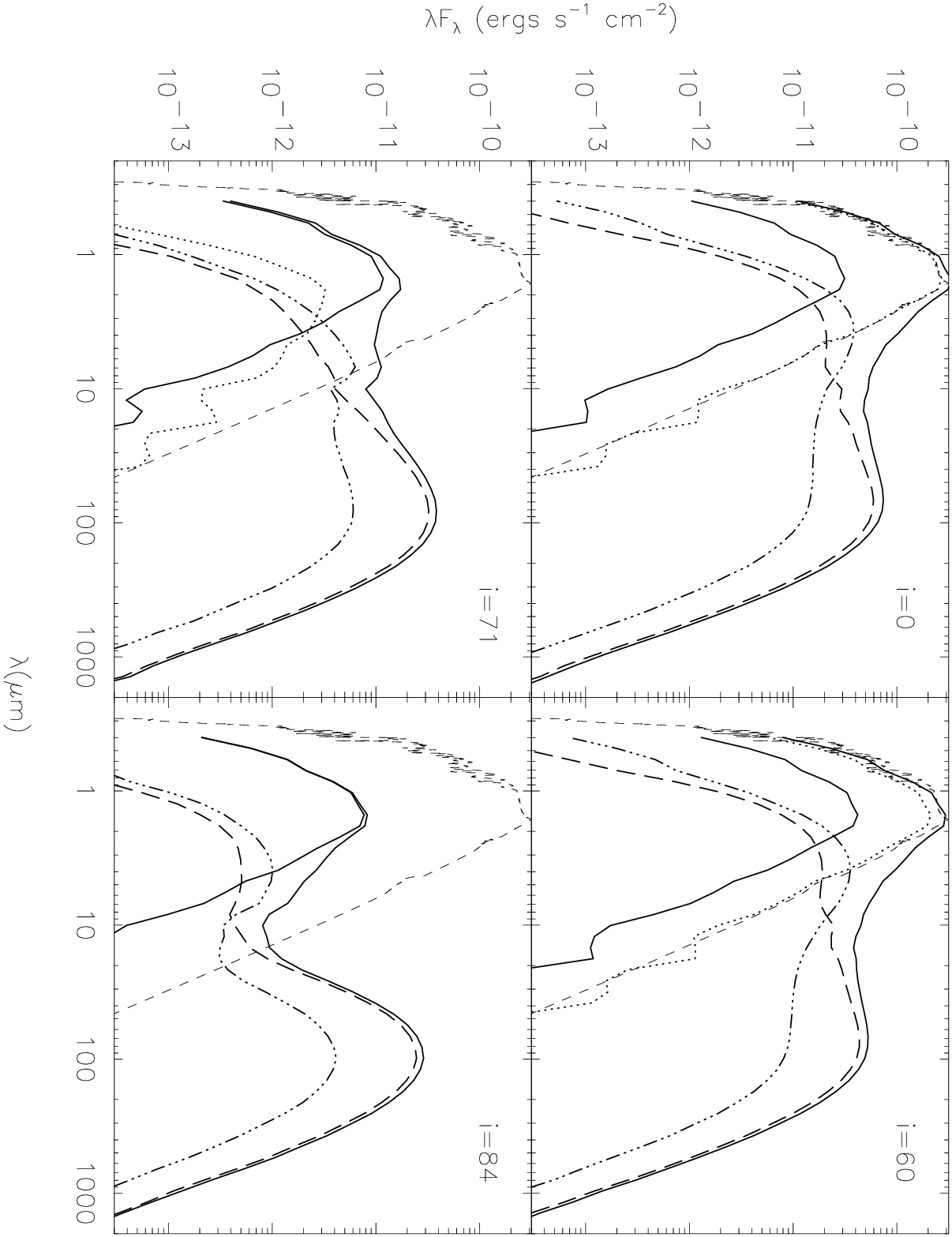}{6in}{90}{65}{65}{-20}{0}}
\caption{Model showing the component parts of the SED for dust Model~1 
at inclinations $i=0^\circ$, 
$i=60^\circ$, $i=71^\circ$, and $i=84^\circ$ (lower panel).  
Each panel shows 
the input stellar spectrum (short dash), total SED (uppermost 
solid line), direct photons (dots), scattered starlight (lowest solid line), 
photons reprocessed by the disk (long dash), and accretion luminosity 
(dash dot dot).  Direct starlight is not detectable for 
$i=84^\circ$ and scattered starlight dominates the 
SED at optical and near-IR wavelengths where the star is totally obscured by 
the flared disk.}
\end{figure}

\end{document}